\documentclass[%
 reprint,
 superscriptaddress,
  amsmath,amssymb,
  aps,
 pra,
]{revtex4-2}
\usepackage{amsmath}
\usepackage{amsfonts}
\usepackage{amssymb}
\usepackage{bm}% bold math

\usepackage{graphicx}

\usepackage{color}

\usepackage[usenames,dvipsnames,svgnames,table]{xcolor}
\usepackage{physics}
\usepackage{quantikz}
\usepackage{hyperref}
\hypersetup{
	colorlinks=true,
	citecolor=ForestGreen,
	linkcolor=ForestGreen,
	urlcolor=ForestGreen,
}

\begin{document}

\title{
Demonstration of a quantum comparator on an ion-trap quantum device
}
% Force line breaks with \\

\author{Tatsuhiko N. Ikeda}
% \thanks{These authors contributed equally to this work.}
\email[email: ]{tatsuhiko.ikeda@riken.jp}
\affiliation{RIKEN Center for Quantum Computing, Wako, Saitama 351-0198, Japan}
\affiliation{BlocQ, Inc., 1-3-1 Kita Aoyama, R Cube Aoyama 3rd Floor, Minato-Ward, Tokyo 107-0061, Japan
}
\affiliation{Faculty of Social Informatics, ZEN University, Zushi, Kanagawa, 249-0007, Japan}
\affiliation{Department of Applied Physics, Hokkaido University, Sapporo, Hokkaido, 060-8628, Japan}

\author{Riku Nakama}
\affiliation{BlocQ, Inc., 1-3-1 Kita Aoyama, R Cube Aoyama 3rd Floor, Minato-Ward, Tokyo 107-0061, Japan
}
\affiliation{Department of Applied Physics, The University of Tokyo, 7-3-1 Hongo, Bunkyo-ku, Tokyo 113-0033, Japan}

\author{Shunsuke Saeki}
\affiliation{BlocQ, Inc., 1-3-1 Kita Aoyama, R Cube Aoyama 3rd Floor, Minato-Ward, Tokyo 107-0061, Japan
}
\affiliation{Department of Applied Physics, The University of Tokyo, 7-3-1 Hongo, Bunkyo-ku, Tokyo 113-0033, Japan}

\author{Hiroki Kuwata}
\affiliation{BlocQ, Inc., 1-3-1 Kita Aoyama, R Cube Aoyama 3rd Floor, Minato-Ward, Tokyo 107-0061, Japan
}
\affiliation{Department of Physics, The University of Tokyo, 7-3-1 Hongo, Bunkyo-ku, Tokyo 113-0033, Japan}

\author{Shuhei M. Yoshida}
\affiliation{BlocQ, Inc., 1-3-1 Kita Aoyama, R Cube Aoyama 3rd Floor, Minato-Ward, Tokyo 107-0061, Japan
}

\author{Akira Shimizu}
\affiliation{BlocQ, Inc., 1-3-1 Kita Aoyama, R Cube Aoyama 3rd Floor, Minato-Ward, Tokyo 107-0061, Japan
}
\affiliation{Institute for Photon Science and Technology, The University of Tokyo, 7-3-1 Hongo, Bunkyo-ku, Tokyo 113-0033, Japan}
\affiliation{Center for Quantum Information and Quantum Biology, The University of Osaka, Toyonaka, Osaka 560-0043, Japan}

%\author{Takashi Nishina}
%\affiliation{BlocQ, Inc., @@@, Japan}

\author{Sho Sugiura}
\email[email: ]{sho.sugiura@blocqinc.com}
\affiliation{BlocQ, Inc., 1-3-1 Kita Aoyama, R Cube Aoyama 3rd Floor, Minato-Ward, Tokyo 107-0061, Japan
}

\date{\today}% It is always \today, today,

\begin{abstract}
Quantum computers are believed to solve a class of computational problems that are based on modular arithmetic faster than classical computers. 
Among the arithmetic building blocks, comparison of integer pairs is a primitive.
Here we report its demonstration in the Reimei quantum computer at RIKEN, whose trapped-ion architecture provides all-to-all qubit connectivity together with high gate fidelities.
We observe high success probabilities
for bit widths $n = 3, 5, 7$, and $9$: 
Under a conventional output-only success criterion we obtain $95\%$ at $n=9$; under a stricter criterion additionally requiring the ancilla to be correct, the success is $69\%$ at $n=9$.
These results demonstrate reliable quantum comparison at scales far beyond those previously achieved experimentally, not only for comparators but also in the broader context of quantum arithmetic circuits.\end{abstract}

\pacs{Valid PACS appear here}% PACS, the Physics and Astronomy
                              %display desireda
                              
\maketitle

%\tableofcontents

%%%%%%%%%%%%%%%%%%%
%{\em Introduction.}---
\section{Introduction}
Quantum computers are expected to provide computational advantages for certain classes of problems~\cite{NielsenChuang} such as integer factorization~\cite{Shor1994}, optimization~\cite{Durr1999}, and quantum computational finance~\cite{Rebentrost2018,Woerner2019,Stamatopoulos2024}. 
These tasks rely heavily on integer and modular arithmetic, including addition, comparison, and modular reduction~\cite{Beauregard2003,Gidney2021}.

Among arithmetic primitives, comparison of integers plays a particularly important role. 
For example, comparators are essential for modular reduction, conditional branching, and decision-making subroutines that appear ubiquitously in quantum algorithms, including Grover-based minimum finding, and quantum optimization routines~\cite{Durr1999,Montanaro2016}. At the same time, comparison circuits are challenging to implement on noisy intermediate-scale quantum (NISQ) devices~\cite{Shahzad2023}, as they require sequential propagation of carry or borrow information and therefore exhibit circuit depths that scale linearly with the bit width~\cite{Cuccaro2004,Takahashi2010,Gidney2018}. As a result, existing experimental demonstrations of quantum arithmetic have largely been restricted to very small problem sizes~\cite{Nishio2020,Li2023, Yalcinkaya2017,Jakhodia2022,Gaur2023,Gaur2024}, 
leaving the regime of scalable comparison largely unexplored.

In this paper, we report an experimental demonstration of comparison
of ordered pairs of $n$-bit integers
on the Reimei quantum computer at RIKEN, whose trapped-ion architecture provides all-to-all qubit connectivity together with high gate fidelities.
Despite the substantial circuit depth, which increases linearly with $n$,
we observe high success probabilities
for bit widths $n = 3, 5, 7$, and $9$: 
Under a conventional output-only success criterion we obtain $95\%$ at $n=9$; under a stricter criterion additionally requiring the ancilla to be correct, 
the success is $69\%$ at $n=9$.
These results establish quantum comparison as a viable arithmetic primitive on current quantum hardware and provide an important benchmark toward realizing digital quantum applications beyond toy arithmetic demonstrations.

%######################################################
\section{Quantum Comparator}\label{sec:comparator}
We consider the problem of comparing two nonnegative integers $a$ and $b$, each represented by an $n$-bit binary string,
\begin{equation}
a = \sum_{i=0}^{n-1} a_i 2^i, \qquad
b = \sum_{i=0}^{n-1} b_i 2^i,
\end{equation}
and determining their ordering relation, i.e., whether $a<b$ or $a\geq b$, in a fully reversible and coherent manner.
In a quantum setting, this task must be implemented by a unitary circuit that maps
\begin{equation}\label{eq:Qcomp}
\ket{a}_n\ket{b}_n\ket{c}_1 \longrightarrow \ket{a}_n\ket{b}_n\ket{c\oplus f(a,b)}_1,
\end{equation}
where $|a\rangle_n=|a_{n-1}\dots a_1a_0\rangle$ and $|b\rangle_n=|b_{n-1}\dots b_1b_0\rangle$ denote $n$-qubit product states.
The Boolean function $f(a,b)$ encodes the comparison result as
\begin{align}
f(a,b)=\begin{cases}
    1 & (a<b)\\
    0 & (a\ge b)
\end{cases}
\end{align}
into the one-bit output register $\ket{c}_1$.
Equation~\eqref{eq:Qcomp} shows only the logical qubits relevant to the comparison; an explicit circuit implementation may require additional ancilla qubits.

\begin{figure}
    \centering
    \includegraphics[width=0.95\linewidth]{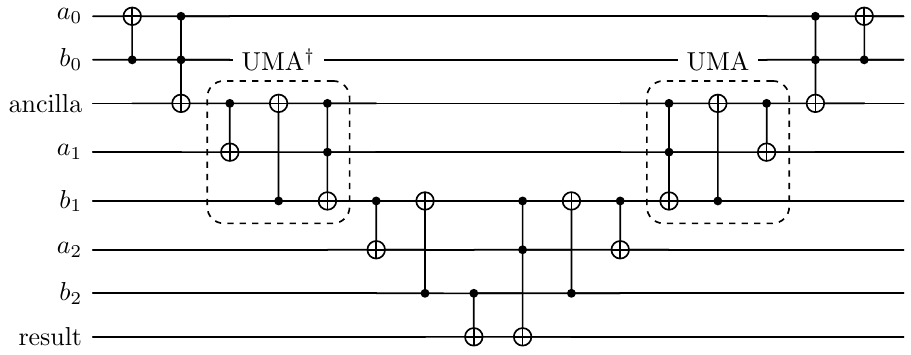}
    \caption{Quantum comparator circuit adopted in this work, exemplified for $n=3$.
    The circuit is adapted from Ref.~\cite{Gouzien2023}.
    The third qubit from the top is a clean ancilla qubit used for borrow propagation, and the total number of qubits is $2n+2$.}
    \label{fig:circuit}
\end{figure}

While various architectures for quantum comparators have been proposed~\cite{Xia2018,Shahzad2023}, in this work we employ an adder-based design originally introduced by Cuccaro et al.~\cite{Cuccaro2004,Gouzien2023}.
This approach is chosen for its minimal ancilla overhead and its decomposability into elementary quantum gates, while maintaining a moderate CNOT counts. Figure~\ref{fig:circuit} shows the comparator circuit used in this work, exemplified for $n=3$.
The circuit is adapted from the construction introduced in Ref.~\cite{Gouzien2023}, and is conceptually equivalent to a reversible subtraction $a-b$, where the sign of the result is extracted without explicitly storing the full difference~\cite{Cuccaro2004}.
Specifically, the comparison outcome is determined by the final borrow bit obtained after sequentially processing the bit pairs $(a_i,b_i)$ from the least significant bit to the most significant bit.

The circuit consists of a cascade of elementary comparator cells, each acting on a single pair of bits and an ancilla qubit that stores the borrow.
These cells update the borrow according to a classical ripple-borrow rule, which can be implemented reversibly using Toffoli and controlled-NOT gates.
As a result, the circuit depth scales linearly with the bit width $n$, making the comparator increasingly sensitive to gate errors as $n$ increases~\cite{Cuccaro2004,Takahashi2010,Gidney2018}.
This linear-depth structure renders the quantum comparator a stronger benchmark for arithmetic operations on NISQ 
% noisy intermediate-scale quantum 
devices.

%######################################################
\section{Experimental Setup}
\begin{figure}
    \centering
    \includegraphics[width=0.95\linewidth]{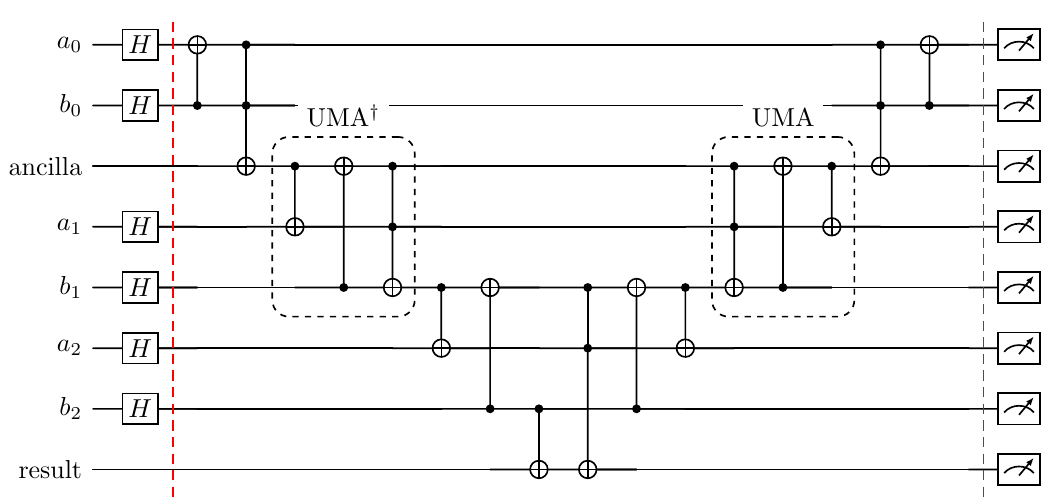}
    \caption{Experimental quantum circuit used in this work, exemplified for $n=3$.
    The circuit consists of preparation of an input state using Hadamard gates, the comparator shown in Fig.~\ref{fig:circuit}, and measurement of all qubits in the computational basis.}
    \label{fig:ExpCircuit}
\end{figure}

All qubits are initialized in the computational basis state $\ket{0}$ at the beginning of each circuit execution.
The experimental quantum circuit used in this work is shown in Fig.~\ref{fig:ExpCircuit} for the representative case of $n=3$.
As illustrated in the figure, the circuit consists of three successive stages:
(i) preparation of an input state using Hadamard gates,
(ii) execution of the quantum comparator circuit, and
(iii) measurement of all qubits in the computational basis.

In the first stage, each qubit in the $n$-qubit registers encoding the integers $a$ and $b$ is acted on by a Hadamard gate.
This prepares the following state
\begin{equation}
\frac{1}{2^n} \sum_{a,b=0}^{2^n-1} \ket{a}_n \ket{b}_n ,
\end{equation}
which is essentially the uniform distribution over all $2^n \times 2^n$ ordered pairs of $n$-bit integers.
In the second stage, the quantum comparator circuit described in Sec.~\ref{sec:comparator} is applied.
This stage implements the comparison operation for each of the input pairs within a single circuit execution.
In the final stage, all qubits, including the input registers, the ancilla qubit, and the output qubit, are measured in the computational basis.

We note that the phase-flip ($Z$) errors (anti)commute with the CNOT and Toffoli gates and do not affect the statistics of measurement outcomes. Thus the phase coherence, even if created in the first stage, might not survive in later stages. The relevant noises are rather the bit-flip ($X$) errors that decrease success rates discussed below.

The circuit uses two $n$-qubit registers for the inputs $a$ and $b$, one clean ancilla qubit for borrow propagation, and one output qubit that stores the comparison result.
The total number of qubits used is therefore $2n+2$.

%######################################################
\section{Results}
%\section{Results and discussions}

We executed the comparison circuit described above 
on the Reimei quantum computer at RIKEN,
which is based on trapped-ion qubits provided by Quantinuum.
For bit widths $n=3,5,7,$ and $9$, 
we performed $100$ shots of experiments for each $n$.
No postselection or error mitigation~\cite{Shaydulin2021} is applied.
The results are summarized in this section.

\subsection{Conventional criterion}\label{sec:conventional}

It is conventional in experimental studies of quantum arithmetic to define the \emph{success count} as the number of measurement outcomes that yield the correct logical comparison result, regardless of the state of ancillary qubits~\cite{Nishio2020,Shahzad2023}.

\begin{figure}
    \includegraphics[width=\linewidth]{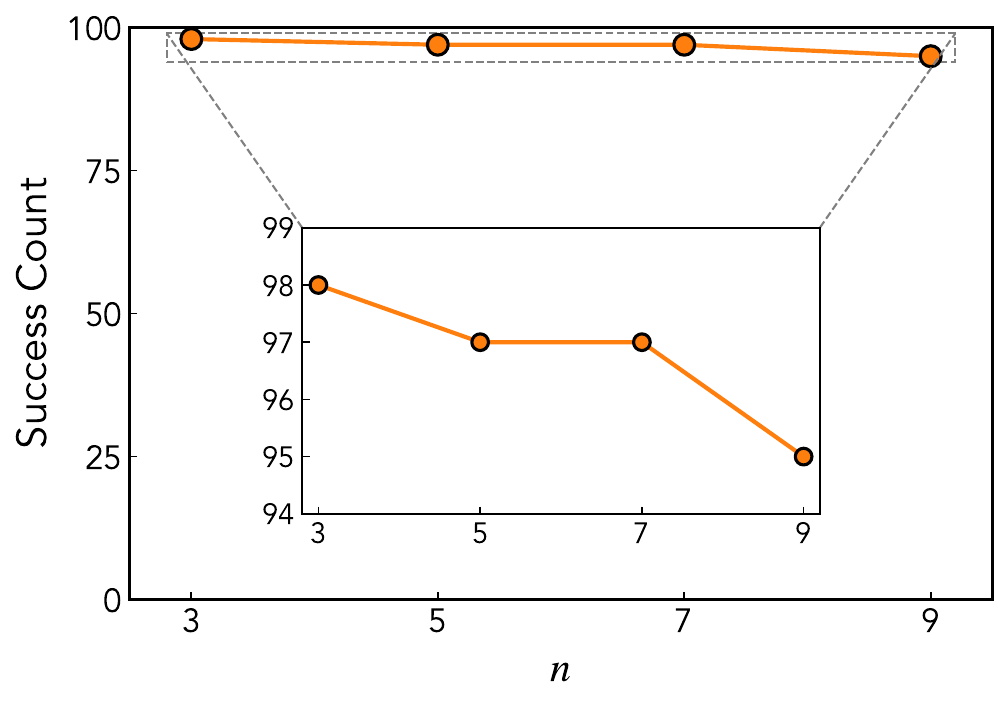}
\caption{The \emph{success count} under the conventional criterion
(see Sec.~\ref{sec:conventional})
is plotted as a function of bit width $n$.
Obtained from $100$ experimental shots for each $n$.
}
\label{fig:results_conv}
\end{figure}

Figure~\ref{fig:results_conv} shows the success probability obtained under this conventional criterion as a function of $n$.
We obtain success probabilities of approximately $98\%$, $97\%$, $97\%$, and $95\%$ for $n=3,5,7,$ and $9$, respectively.
These values are significantly higher than the success probability of random outputs, which is $50\%$.
The observed high success probabilities significantly exceed those in prior experimental demonstrations of a quantum comparator on real hardware, which were restricted to $n=2$ and classical inputs~\cite{Shahzad2023}.

The same result can be expressed in terms of the failure probability, defined as one minus the success probability.
For random outputs, the failure probability is also $50\%$.
In contrast, our measured failure probabilities are approximately $2\%$, $3\%$, $3\%$, and $5\%$ for $n=3,5,7,$ and $9$, respectively.
Thus, even at the largest bit width studied, we achieve more than an order-of-magnitude reduction in the failure probability compared with random outputs.

\subsection{Stricter criterion}\label{sec:strict}

To gain further insight into the error mechanisms, we analyze the experimental outcomes by classifying them into four categories:
(i) \emph{ancilla-inclusive success}, where both the comparison result and the ancilla qubit are correct;
(ii) \emph{fail:~$a<b$ result}, where the logical comparison outcome is incorrect;
(iii) \emph{fail:~ancilla}, where the logical comparison result is correct but the ancilla qubit is flipped; and
(iv) \emph{fail:~both}, where both the comparison result and the ancilla qubit are incorrect.

The category \emph{ancilla-inclusive success} requires not only the correct logical comparison result but also the absence of any error in the ancilla qubit, and therefore constitutes a stricter criterion for successful computation.
By contrast, the conventional \emph{success count} discussed in Sec.~\ref{sec:conventional} corresponds to the sum of \emph{ancilla-inclusive success} and \emph{fail:~ancilla} events.

\begin{figure*}
    \includegraphics[width=\linewidth]{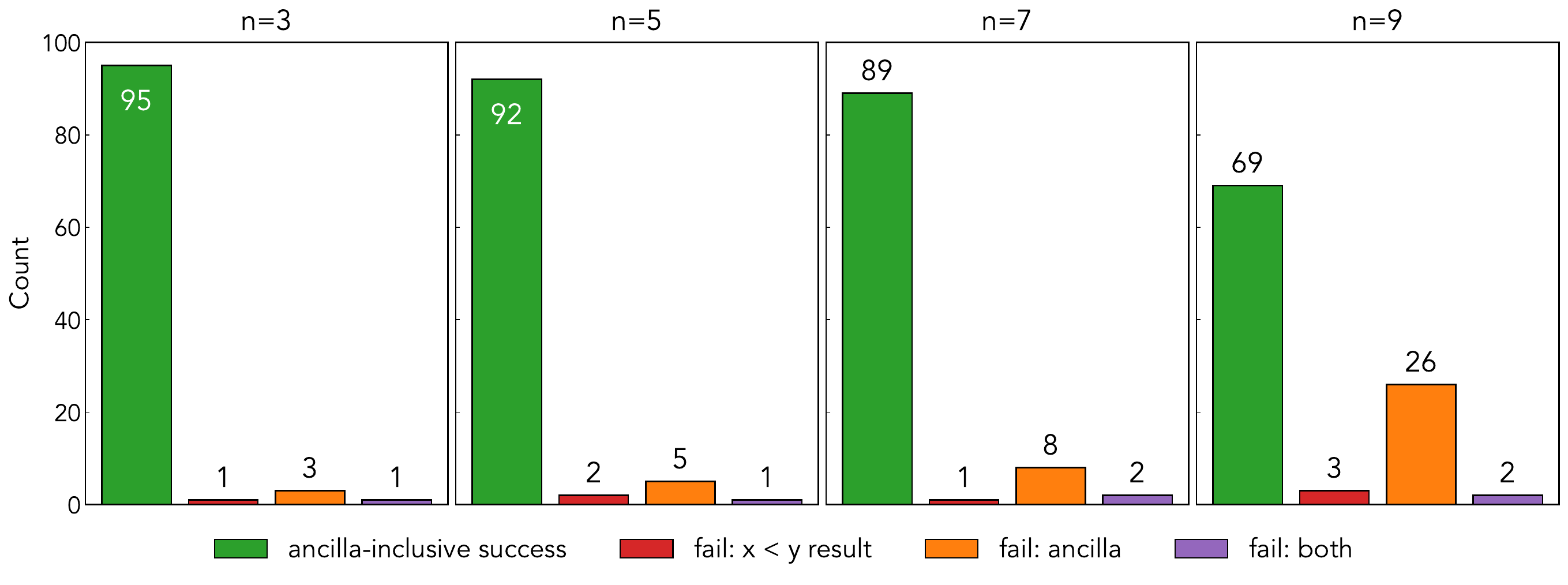}
    \caption{Histograms of the four outcome categories 
    introduce as a stricter criterion
in Sec.~\ref{sec:strict}. 
Obtained from $100$ experimental shots for each bit width $n$.}
    \label{fig:results}
\end{figure*}

Figure~\ref{fig:results} shows histograms of the four categories for each value of $n$.
Even under this stricter criterion, the probability of \emph{ancilla-inclusive success} remains high, taking values of approximately $95\%$, $92\%$, $89\%$, and $69\%$ for $n=3,5,7,$ and $9$, respectively.
For comparison, the probability of \emph{ancilla-inclusive success} for random outputs is $25\%$.
Thus, even at $n=9$, we achieve a probability of \emph{ancilla-inclusive success} that is higher than the random baseline by a factor of $69\%/29\%\approx3$.

A notable feature of the data is that the dominant failure mode at larger $n$ arises from ancilla-only errors.
This behavior can be naturally understood from the structure of the comparator circuit.
The ancilla qubit stores the borrow information throughout the entire depth of the circuit and is repeatedly involved in multi-qubit controlled operations as a logical workspace, making it particularly susceptible to the accumulation of gate and decoherence errors~\cite{Cuccaro2004,Gidney2018}.
In contrast, the logical comparison result is transferred to the output qubit only at the final stage of the circuit, rendering it comparatively robust.
%\cmtas{Is the following statement applicable to circuits that have no ancillas? Should we romove it?}
%\as{The observed error pattern therefore reflects an intrinsic property of linear-depth reversible arithmetic circuits rather than a flaw specific to the present implementation~\cite{Cuccaro2004}.}

\section{Discussions}\label{sec:discussion}
Whether ancilla-only errors should be tolerated depends on the intended application.
When the comparator is used as a standalone arithmetic operation, the conventional criterion based on the \emph{success count} is reasonable, since the logical comparison result remains correct even if the ancilla qubit is flipped.
However, when the comparator is embedded as a subroutine in a larger quantum circuit, ancilla errors may propagate and corrupt subsequent operations.
In such contexts, the stricter criterion based on \emph{ancilla-inclusive success} provides the appropriate performance metric.

Importantly, even under this stricter criterion, our results demonstrate correct quantum comparison at bit widths far beyond those previously reported in experimental studies.
The success probabilities remain high despite the use of the linear-depth structure of the circuit, underscoring the robustness of the logical comparison operation itself.

Overall, these findings establish that comparison can be executed reliably on current quantum hardware at scales significantly larger than prior experimental benchmarks.
The observed error patterns further suggest that improvements in ancilla fidelity, rather than in the logical comparison operation, are likely to yield substantial gains in performance at larger bit widths.

A plausible explanation for the strong performance observed in our experiments is the hardware characteristics of the ion-trap quantum processor~\cite{Wright2019,Bruzewicz2019,Georgescu2020} used in this work.
In particular, the combination of high two-qubit gate fidelities and all-to-all qubit connectivity may be especially well suited for arithmetic circuits with frequent multi-qubit interactions, such as the quantum comparator studied here.

\section{Summary and Outlook}
In this work, we have experimentally demonstrated a comparison circuit of two $n$-bit integers on real quantum hardware for bit widths up to $n=9$.
Under the conventional success criterion, which requires only the correct logical comparison result, we observe success probabilities of approximately $98\%$, $97\%$, $97\%$, and $95\%$ for $n=3,5,7,$ and $9$, respectively.
Under the stronger success criterion, which additionally requires the absence of any ancilla error, the corresponding success probabilities are approximately $95\%$, $92\%$, $89\%$, and $69\%$.
To our knowledge, these results represent the largest-scale experimental realization of a quantum arithmetic comparison circuit,
and they significantly exceed the success probabilities reported in prior experimental studies.

As discussed in Sec.~\ref{sec:discussion}, the appropriate interpretation of the success probability depends on the intended use of the comparator.
Nevertheless, even under the stricter criterion, our results demonstrate 
a substantially higher success probability for quantum comparison than that in previous studies, as well as a success probability that is substantially higher than that of random sampling in a regime that remains experimentally unexplored for quantum arithmetic circuits. For further evaluation of our demonstration, an estimation of the phase is anticipated. 

Looking ahead, improvements of the circuit are expected from techniques such as implementation with less depth~\cite{Remaud_2024} and usage of relative phase Toffoli gates~\cite{PhysRevA.52.3457,PhysRevA.93.022311}.
More broadly, we anticipate that quantum comparators will serve as a valuable benchmark for assessing the readiness of quantum processors for digital quantum applications and for building larger arithmetic circuits that underpin quantum algorithms based on modular arithmetic.

\section*{Author Contributions}
T.N.I.\ conceived the project and designed the experimental setup.
T.N.I., R.N., and S.\,Saeki constructed the comparator circuits.
S.M.Y.\ provided technical guidance on circuit compilation and
optimization using the native gates of the device.
T.N.I.\ compiled and executed the circuits on the Reimei quantum computer
at RIKEN and acquired the data.
T.N.I., R.N., S.\,Saeki, and H.K.\ analyzed the experimental results and
prepared the figures.
A.S.\ contributed substantially to the interpretation of the results,
including the analysis of the error mechanisms and the success criteria,
as well as to the writing of the manuscript.
S.\,Sugiura supervised the project and coordinated the collaboration among
the participating institutions.
T.N.I., and A.S. wrote the manuscript with substantial feedback from the other authors.
All authors discussed the results and contributed to the project.

\begin{acknowledgments}
This work was
supported in part by the New Energy and Industrial Technology Development Organization (NEDO), Japan (Project No.
JPNP20017).
T.~N.~I. was supported by Japan Society for the Promotion of Science (JSPS) KAKENHI Grant No.~25K07178.
A.S. was supported by JPSJ
KAKENHI Grant No. 23K22413 and by the RIKEN TRIP initiative.
\end{acknowledgments}

\section*{Data Availability}
The data that support the findings of this study, including the compiled circuit descriptions at the native-gate level, are available from the corresponding author upon reasonable request.

\bibliography{blocqreferences}% Produces the bibliography via BibTeX.

\end{document}